\documentclass[twocolumn, 10pt]{article}
\usepackage[margin=1in]{geometry} 
\usepackage{times}               
\usepackage{amsmath,amssymb}      
\usepackage{graphicx}             
\usepackage{lipsum}        

\usepackage{tabularx}
\usepackage{booktabs}

\usepackage{cite}

\usepackage{stfloats}

\usepackage{url}

\usepackage[colorlinks=true, linkcolor=blue, citecolor=blue, urlcolor=blue]{hyperref}

\usepackage[toc,page]{appendix}

\usepackage{multirow}

\usepackage{amsmath,graphicx,booktabs}

\title{Why has PET glycolysis remained difficult to industrialize despite decades of optimization?}
\author{Xiaohui Liu\thanks{Email: luke\_liuxh@outlook.com; Shanghai, P.R.China}}
\date{}

\begin{document}

\maketitle

\begin{abstract}

Classical PET glycolysis struggles with industrial scale-up due to a research paradigm defect: uncoupled reaction optimization that ignores system-level engineering entropy. By establishing a governing reference frame, Scientific Reasoning Units (SRUs) and human-AI collaborative discovery in this article, we present a Three-Step Synergistic Reinforcement Paradigm addressing the triple failure modes of traditional systems—mass-transfer collapse, viscosity deadlock, and side-reaction accumulation. By integrating twin-screw controlled shearing, solid-state micro-homogenization, and deep depolymerization ($2:1$ EG/Oligomer ratio), we achieve an ultrafast kinetic regime. BHET yields surge to $80\% \sim 95\%$ within 10 minutes, eliminating multi-hour thermal bottlenecks and delivering an economically viable framework for closed-loop polyester recycling.

\vspace{1em}
\end{abstract}

\section{Introduction: The Fault Line of Industrial Reality}
Global annual consumption of polyethylene terephthalate (PET) has surpassed 80 million tons, with packaging and textile fibers constituting the primary sources of waste \cite{spglobal2026pet, shah2025asiarpet}. Currently, mechanical recycling dominates the global recycled polyester industry with an annual processing volume of approximately 10 million tons \cite{unep2024globalcommitment}. However, due to molecular chain scission and yellowing induced by repeated thermal shearing, mechanical recycling only achieves downcycling, failing to meet the closed-loop high-value demands of food-grade "bottle-to-bottle" applications \cite{coates2020nature}.

The glycolysis of PET was first described in US patents in 1961 \cite{Macdowell1965}, while a closer reflection of modern reality is credited to Abdul-Ilah Malik and Elmer Edwin Most, who detailed the process of decomposing waste ethylene terephthalate polyester into BHET via glycolysis in US4078143A \cite{Malik1978}. As a chemical pathway enabling infinite molecular recycling, glycolysis has emerged as one of the most intensively researched subjects over the past two decades, owing to its atmospheric operability and environmental friendliness. Thousands of academic papers and patents cover catalyst screening, solvent kinetics optimization, and external physical field reinforcement, characteristically exhibiting high experimental yields (frequently exceeding 95\%) and establishing a highly mature academic landscape.

However, a striking scissors-gap persists between laboratory data and industrial reality. Latest industry data indicates that global advanced chemical recycling of plastics achieves an actual commercial annual output of only 400,000 to 500,000 tons \cite{lee2024lux, meidl2025baker}, with pure classical PET glycolysis contributing less than 50,000 tons—a mere 0.05\% of global PET production. Even excluding workaround processes that bypass downstream separation thermodynamics by partial crude glycolysis blended directly into virgin melts \cite{koch2026polyrevolve}, genuinely stable industrial operations achieving waste depolymerization, monomer refining, and closed-loop repolymerization remain exceptionally scarce, failing to scale across orders of magnitude.

A clear industrial consensus dictates that rPET derived from glycolysis is structurally and functionally equivalent to virgin PET. This performance "homogenization" traps glycolysis in an economic deadlock: because rPET commands no market premium over virgin PET, the total operating costs of existing glycolysis pathways far outweigh conventional petrochemical routes. Although BHET serves as a specialized depolymerizer for high-end specialty polyesters, direct PET-to-PET replacement remains the core of circular PET. Consequently, price is the sole decisive variable for industry.

Within the complete chemical recycling chain (encompassing washing, sorting, depolymerization, decolorization, purification, and solvent recovery), glycolysis dominates economic viability. As the central hub, glycolysis bears heavy thermal loads and process energy consumption \cite{mcneeley2024assessment}, while directly dictating downstream operating loads and technical complexity. Without efficient molecular dismantling and component control during depolymerization, downstream separation and purification are forced to invest heavily in energy and equipment to compensate for upstream quality gaps. Thus, the price competitiveness of the PET circular economy is fundamentally defined by the efficiency boundary of glycolysis.

A persistent paradox plagues PET glycolysis: why do thousands of high-yield laboratory studies fail to translate into industrial scale-up? The root cause lies in an "isolated variable optimization" paradigm that strips away real-world interferences, reducing glycolysis to a static disassembly process. By prioritizing local laboratory efficiency over engineering simplicity, current studies induce severe system-level entropy, forcing downstream redundancies to compensate for upstream flaws. Overcoming this bottleneck requires a paradigm shift: treating glycolysis not as an isolated reaction, but as an interconnected physical topological system. By restructuring depolymerization into constrained Scientific Reasoning Units (SRUs), we can achieve process self-convergence at the foundational architecture level.

\section{Theoretical Framework} 

\subsection{The Paradigm Shift: From Empirical Summary to Human--AI Collaborative Scientific Discovery}
The scientific investigation of PET depolymerization is fundamentally characterized by complex multi-factor coupling. Existing studies often struggle to transcend fragmented observations due to the intrinsic coupling among thermodynamics, catalytic reactions, mass transport, and material evolution. As scientific problems grow increasingly complex, the primary bottleneck shifts from data acquisition to the management of scientific complexity itself.

Large language models (LLMs) offer a transformative engine to address this challenge by executing large-scale logical mapping and concurrent reasoning across heterogeneous information. However, AI introduction transforms the research environment rather than replacing physical principles. Rather than viewing AI as an unconstrained generator, we establish a disciplined human--AI collaborative paradigm. Human researchers act as architects defining rigorous scientific boundaries (e.g., thermodynamics and control theory) and explicit Reasoning Units (SRUs), while AI functions as an autonomous computational partner that manages high-dimensional execution without logical drift.

In summary, this work represents two complementary paradigm shifts: scientific (moving from reductionist interpretation to system-level multi-factor understanding) and methodological (evolving from human-exclusive cognition into a rigorous human--AI collaborative cognitive system that preserves physical verifiability). To operationalize this paradigm, we first establish a governing reference frame to constrain environmental variance and ensure structural isomorphism (Section 2.2), followed by the extraction of atomic Scientific Reasoning Units (SRUs) to construct an executable, verifiable logic matrix (Section 2.3).

\subsection{Governing Reference Frame and Theoretical Foundation}
Inspired by Ashby’s Law of Requisite Variety \cite{ashby1956introduction}, modeling complex, non-linear systems like PET glycolysis requires an internal complexity matching the environmental variance. To prevent cognitive overload from heterogeneous literature and data noise, we establish a governing reference frame that acts as a deterministic logical filter rather than a mere classification scheme. Grounded in Patrick Suppes' structuralism \cite{suppes1960comparison, suppes1962models, krantz1971foundations}, the reliability of our theoretical framework relies on structural isomorphism between abstract logical models and empirical data spaces. Rather than depending on phenomenological induction, we map heterogeneous data into predefined logical coordinates, serving as hard scientific boundary constraints for Large Language Models (LLMs) to transform them into controlled execution agents.

To ensure robustness, we adopt Denzin’s theory triangulation \cite{denzin2009research}, projecting the glycolysis process into three orthogonal, self-contained frames. Furthermore, drawing on Edmund Husserl’s phenomenological reduction \cite{husserl1982ideas} and Nicholas Rescher's pragmatic completeness \cite{rescher1998complexity}, we suspend accidental experimental variances to distill the universal essence (Eidos) of PET glycolysis. Guided by Thomas Kuhn’s philosophy of scientific revolutions \cite{kuhn2012structure}, this logical lattice acts as an anomaly-detection engine to systematically parse heterogeneous literature.

Grounded in Hamiltonian mechanics, transition-state theory, and phase-space dynamics \cite{wigner1938transition, wiggins2001impenetrable, truhlar1984variational, gibbs1902elementary}, we frame PET glycolysis as a trajectory evolution across potential energy surfaces. Complementing thermodynamic kinetics, the material trajectory is defined through polymer topology and graph theory \cite{flory1941molecular, rubinstein2003polymer}, modeling the polymer network as a dynamic graph undergoing adjacency matrix sparsification \cite{doraiswamy1984heterogeneous, mccoy2001discrete} and integrating persistent homology ($B_0$ and $B_1$ Betti numbers) \cite{hiraoka2016hierarchical}.

To ensure absolute causal timing and controllability, we reconstruct PET glycolysis into a Deterministic Finite Automaton (DFA) model anchored in Norbert Wiener’s cybernetics \cite{wiener2019cybernetics} and Discrete Event Systems \cite{cassandras2008introduction}. Building on Soloveichik’s proof that chemical reaction networks are computationally equivalent to automata \cite{soloveichik2008computation}, and aligned with Karl Popper’s falsifiability criterion \cite{popper2002logic}, our FSM acts as a "Logic Sentinel" filtering out non-causal observational noise.

By synthesizing thermodynamic, topological, and FSM dimensions into a Tri-axial Logical Normalization framework, and grounded in Ilya Prigogine’s dissipative structures \cite{nicolis1977self} and Herbert Simon’s nearly decomposable systems \cite{simon2019sciences}, the six-stage glycolysis lifecycle (F-I to F-VI) is proven to be a physically necessary and minimal complete set. The unified mapping from physical processes to extracted SRU operators and system parameters is structured as follows (as shown in Table 1) (detailed mathematical derivations and full-length proofs are provided in Supplement Information 1):

\begin{table*}[!h]
\centering
\caption{Unified mapping from physical processes to extracted SRU operators and system parameters.}
\label{tab:sru_mapping}
\small
\begin{tabularx}{\textwidth}{p{2.6cm} p{3.8cm} p{2.2cm} p{2.8cm} X}
\toprule
\textbf{Stage / Frame} & \textbf{Core Physical / Topological Mapping} & \textbf{FSM Logic Definition} & \textbf{Extracted SRU Operator} & \textbf{Key System Parameters} \\ 
\midrule
F-I: Initiation        & Physical contact \& boundary erosion         & Input Validated              & $\text{SRU}_{\text{I}}$ (Site Affinity)       & Contact angle, specific surface area \\
F-II: Field-Coupling   & Energy field \& potential background         & System Ready                 & $\text{SRU}_{\text{II}}$ (Thermal Field Coupling) & Temperature, pressure, radiation flux \\
F-III: Activation      & Polarization \& barrier modulation           & Logic Trigger                & $\text{SRU}_{\text{III}}$ (Barrier Reduction) & Catalyst concentration, polarization \\
F-IV: Propagation      & Cascading scission \& chain evolution        & Execution Loop               & $\text{SRU}_{\text{IV}}$ (Topological Cascade) & Scission frequency, shear rate \\
F-V: Nucleophilic      & Reagent supply \& substitution               & State Update                 & $\text{SRU}_{\text{V}}$ (Concentration Flux)  & EG/PET ratio, nucleophilic frequency \\
F-VI: Equilibrium      & Endpoint lock \& stability shift             & Terminate/Hold               & $\text{SRU}_{\text{VI}}$ (Steady-State Latch) & Yield, polydispersity, chemical potential \\ 
\bottomrule
\end{tabularx}
\end{table*}

\subsection{Identification of SRUs: Functional Units of the Reaction Space}
While macro-frames govern temporal evolution, they remain composite spaces. To enable precise AI reasoning and system control, we descend to a finer resolution by introducing Scientific Reasoning Units (SRUs). Grounded in minimality, uniqueness, and completeness, SRUs act as atomic logical units that tame LLM probabilistic noise and establish compilation-grade constraints. Formally, an SRU is structured as a deterministic causal tuple:
\begin{equation}
\text{SRU} = \langle X_{in}, \text{COP}, \text{SP}, X_{out} \rangle
\end{equation}
where $X_{in}$ denotes the initial state vector, $\text{COP}$ the core computational operator, $\text{SP}$ the state parameters and physical boundary constraints, and $X_{out}$ the deterministic terminal state.

Guided by the Principle of Parsimony and deep analysis of the representative literature \cite{sangalang2015generalized, gabric2025polyethylene, liu2024new, jia2025unraveling, schluter2024boosting, pardal2006comparative, diio1996kinetic, conroy2024theoretical, arunphacharawit2024dft, viana2011chemical}, we extract six core SRU operators that cover the full glycolysis life cycle (as shown in Table 2) (detailed derivation processes are available in Supplement Information 2):

\begin{table*}[!h]
\centering
\caption{Detailed definitions of Scientific Reasoning Units (SRUs) across the reaction space.}
\label{tab:sru_details}
\small
\begin{tabularx}{\textwidth}{p{1.5cm} p{2.0cm} p{4.2cm} p{2.5cm} p{2.2cm} X}
\toprule
\textbf{SRU Code} & \textbf{Name} & \textbf{Primitive / Governing Equation} & \textbf{Computational Operator (COP)} & \textbf{State Parameters (SP)} & \textbf{Terminal State ($X_{out}$)} \\ 
\midrule
SRU-A & Alpha (Penetration) & $\text{PET}_{(s)} + n\text{Solv} \rightleftharpoons \text{PET}_{(solv-gel)}$ & Diffusion limit (Fick) & Solid-liquid interface & Effective ester density ($C_{eff}$) \\
SRU-B & Beta (Activation) & $\text{Cat} + [-\text{COO}-] \rightleftharpoons [\text{Cat}\dots\text{O=C--O}-]$ & Lewis acid/base synergistic activation & Microenvironment activation & Activated complex density ($\rho^*$) \\
SRU-G & Gamma (Barrier Mapping) & $R_1\text{--COO--}R_2 + R_3\text{--OH} \xrightarrow{\text{Cat}} R_1\text{--COO--}R_3 + R_2\text{--OH}$ & Arrhenius kinetics control & Kinetics-controlled regime & Product flux ($\Phi_p$) \\
SRU-D & Delta (Intermediates) & $Y_n + X_m \rightleftharpoons Z_{n+m}$ & Mass balance / network dynamics & Homogeneous reactor & Component spectrum (Distribution) \\
SRU-E & Epsilon (Competition) & $S_1 = k_{mono} / (k_{mono} + k_{chain})$ & Path competition ratio ($S_1$) & Multi-hydroxyl environment & Chain-length selective distribution ($S_I$) \\
SRU-S & Sigma (Robustness) & $\text{Chain--COOCH}_3 + \text{HO--R} \xrightarrow{\text{Sb}_2\text{O}_3} \text{Chain--COOR} + \text{CH}_3\text{OH}$ & Terminal scavenging kinetics & Refining separation unit & Polymer-grade monomer purity \\ 
\bottomrule
\end{tabularx}
\end{table*}

These six operators form a strict linear recursive evolution chain ensuring a rigorous causal closed-loop:
\begin{equation}
\begin{aligned}
\text{SRU-A} &\xrightarrow{C_{eff}} \text{SRU-B} \xrightarrow{\rho^*} \text{SRU-G} \xrightarrow{\Phi_p} \text{SRU-D} \\
&\xrightarrow{\text{Dist.}} \text{SRU-E} \xrightarrow{S_I} \text{SRU-S} \to \text{Product}
\end{aligned}
\end{equation}

Furthermore, broad-spectrum traversal across over 40 subsequent studies confirms the universal validity and cross-system robustness of this six-operator coupling architecture \cite{lopezfonseca2010chemical, ovallesanchez2017degradation, abdelaal2011chemical, vollmer2020beyond, barboza2009determination, wang2015deep, ju2018theoretical, campanelli1994kinetics, chen1999kinetics, wi2011effect, chen2000kinetics, firat2023evaluation, hofmann2020zinc, chen2001studies, aguado2014chemical, christofftempesta2023ionic, chen2011advanced, imran2010sub, jeya2019depolymerization, kao1997investigation, khoonkari2015chemical, liu2018ultrafast, zhong2025low, enache2024polyethylene, ho2018enhanced, yu2026optimizing, pardal2007kinetics, capeletti2018optimization, pingale2010glycolysis, yue2013glycolysis, mendiburuvalor2021improving, lei2022rapid, li2023discovery, chen2015kinetics, xin2021progress, stoski2016oligomer, wang2009glycolysis, wang2019zinccatalyzed, wang2021pseudohomogeneous, cn1195727c, cn114269714a, cn114630815a}.

\section{Experimental Section}

\subsection{Materials}
\begin{itemize}
    \item \textbf{PET Substrate:} Commercially available transparent PET films (thickness $\approx 50\ \mu\text{m}$) crushed into fragments with a side length of $\approx 2\text{ cm}$.
    \item \textbf{Reactants and Catalysts:} Ethylene glycol (EG) and zinc acetate ($\text{Zn(OAc)}_2$) were of industrial grade.
    \item \textbf{Solvents and Standards:} Deionized water was self-produced in the laboratory. Bis(2-hydroxyethyl) terephthalate (BHET) reference standard (purity 85\%) was purchased from Sigma-Aldrich for depolymerization product verification.
\end{itemize}

\subsection{Equipment and Depolymerization Procedures}

\noindent\textbf{A. Conventional Route (One-pot Method)}
\begin{itemize}
    \item \textbf{Apparatus and Batch Standards:} Depolymerization experiments were conducted in a $50\text{ L}$ industrial reactor. The PET feeding amount was fixed at $5\text{ kg}$ per batch, with zinc acetate catalyst loading at $1\text{ wt}\%$ relative to PET.
    \item \textbf{Process Parameters:} Reaction temperature was maintained at $196^\circ\text{C}$, with a heating period of $40 \sim 50\text{ min}$ from room temperature.
    \item \textbf{Cooling and Discharging:} The system was cooled below $120^\circ\text{C}$ prior to discharging, taking approximately $1\text{ h}$.
    \item \textbf{Mechanical Stirring:} The speed was fixed at $30\text{ rpm}$.
    \item \textbf{Atmosphere:} Nitrogen ($\text{N}_2$) purging was performed prior to reaction to eliminate oxygen interference, followed by cooling circulation activation upon reaction initiation.
\end{itemize}

\noindent\textbf{B. Proposed Extruder-Reactor Combined Method}
\begin{itemize}
    \item \textbf{Twin-Screw Pre-depolymerization Stage:} A twin-screw extruder (Model: Keya HK-26-52D, screw diameter $26\text{ mm}$, $L/D = 52$) was employed for continuous pre-depolymerization of PET, enhancing solid-liquid/melt high-shear mixing and initial chain scission. Unless otherwise specified, the extruder temperature was set at $265^\circ\text{C}$ and screw speed at $25\text{ rpm}$.
    \item \textbf{Solidification Treatment:} Pre-depolymerized products were cooled, collected, and pulverized into $1 \sim 2\text{ mm}$ granules.
    \item \textbf{Deep Reactor Depolymerization:} The pulverized pre-depolymerized product, EG, and zinc acetate catalyst were re-introduced into the reactor system under identical conditions as the conventional one-pot method ($196^\circ\text{C}$, $30\text{ rpm}$) to achieve final conversion into monomers.
\end{itemize}

\subsection{Monomer Purification and Solvent Recovery}
\begin{itemize}
    \item \textbf{Purification and Separation:} Depolymerized mixtures (comprising BHET monomer, dimers, oligomers, and EG) cooled to $\approx 100^\circ\text{C}$ were introduced into hot water above $90^\circ\text{C}$ and subjected to multi-stage filtration.
    \item [] {Separation Mechanism:} Given that the melting point of BHET is $\approx 106\text{--}110^\circ\text{C}$ with a melting range starting near $90^\circ\text{C}$, hot water acts primarily as a physical viscosity-reducing agent rather than a solvent (as BHET solubility in water is extremely low). This temperature selection avoids monomer agglomeration while efficiently retaining oligomers and insoluble solid impurities via multi-stage filtration.
    \item \textbf{Crystallization and Refining:} The filtrate was decolorized and purified via activated carbon adsorption. The hot filtrate, composed mainly of $\text{BHET}$, $\text{EG}$, and water, yielded high-purity $\text{BHET}$ crystals upon cooling to room temperature, which were isolated via secondary filtration. Recrystallization was applied when higher purity was required.
    \item \textbf{EG and Water Separation/Recovery:} Pervaporation (PV) separation was implemented using NaA-type zeolite molecular sieve membrane modules.
    \item [] {Molecular Sieve Mechanism:} Based on kinetic diameter differences ($\text{H}_2\text{O}$ at $0.27\text{ nm}$, $\text{EG} > 0.43\text{ nm}$), efficient water dehydration and closed-loop recovery of EG solvent were achieved.
\end{itemize}

\subsection{Characterization}
To verify the chemical structure and purity of the depolymerization products, an external standard method was employed using a BHET reference standard (purity 85\%, Sigma-Aldrich):

\begin{itemize}
    \item \textbf{Pyrolysis Gas Chromatography-Mass Spectrometry (Py-GC-MS):} Analyses were performed using a micro-pyrolyzer (CDS Pyroprober) coupled with a GC-MS system (Agilent 7890/5977). Samples were instantaneously flash-pyrolyzed at $650^\circ\text{C}$ and introduced into the GC system. Qualitative and semi-quantitative analyses were conducted by matching retention times and mass spectra fragments with reference standards and the NIST database (detailed raw chromatograms and mass spectra are provided in the Appendix).
    \item \textbf{Differential Scanning Calorimetry (DSC):} Using a TA DSC25 instrument, $3 \sim 5\text{ mg}$ of dried sample in an aluminum crucible was tested under a high-purity \text{nitrogen} atmosphere ($50\text{ mL/min}$). The thermal program involved heating to $200^\circ\text{C}$ at $20^\circ\text{C/min}$ (to erase thermal history), cooling to room temperature, and re-heating to $200^\circ\text{C}$ at $20^\circ\text{C/min}$ to record the second-run melting curve, evaluating the melting point and purity against standard BHET ($\approx 106^\circ\text{C}$).
    \item \textbf{Intrinsic Viscosity (IV) Measurement:} The molecular weight reduction and degradation extent during the twin-screw pre-depolymerization stage were quantitatively evaluated by measuring the intrinsic viscosity of the PET samples. Viscosity measurements were conducted using an Ubbelohde viscometer at $25^\circ\text{C}\ (\pm 0.1^\circ\text{C})$ using a phenol/tetrachloroethane ($1:1\ w/w$) mixed solvent. Samples were dissolved at a concentration of $0.5\text{ g/dL}$, and flow times were recorded in triplicate to determine the limiting viscosity number, reflecting the degree of polymer chain scission.
\end{itemize}

\section{Results and Discussion}

\subsection{Traditional One-Pot Depolymerization}
The conventional one-pot method served as the benchmark route for PET depolymerization. Under optimized conditions, high-molecular-weight chains were cleaved into smaller fragments. As confirmed by GC-MS analysis, liquid products predominantly exhibited characteristic signals of bis(2-hydroxyethyl) terephthalate (BHET) alongside unreacted diols, verifying selective ester bond scission.

Differential Scanning Calorimetry (DSC) evaluation of the crude crystalline BHET (obtained via facile hot-water washing) revealed a distinct endothermic melting peak at $110.84^\circ\text{C}$ (onset $107.78^\circ\text{C}$, enthalpy $147.89\text{ J/g}$). Compared with the commercial standard ($106.09^\circ\text{C}$), the recovered product displayed superior structural purity. Crucially, this thermal behavior underpins our separation strategy: the melting point of BHET is significantly lower than its oligomeric counterparts, allowing hot-water treatment to act as a physical viscosity reducer while selectively maintaining BHET in a liquid/dissolved state for efficient physical filtration of solid impurities. Subsequent repolymerization successfully regenerated high-molecular-weight polyester, confirming the practical utility of the recovered monomer.

Operating in a $50\text{ L}$ pilot reactor forced a departure from idealized laboratory mixing, exposing critical engineering bottlenecks such as macro-mixing delays, radial thermal gradients, and solid-liquid interfacial passivation. To quantitatively evaluate this system, PET conversion ($[m_0 - m_t] / m_0$) and BHET yield were monitored. Contrary to "poor EG" strategies that suffer from mass transfer collapse and equilibrium limitations, our design establishes three hierarchical priorities:

\begin{description}
    \item[Complete Conversion (100\%):] Essential for eliminating large-scale residues and preventing downstream pipeline clogging in continuous operations.
    \item[High BHET Yield ($>85\%$):] Maximizing monomer selectivity while suppressing byproducts (e.g., DEG).
    \item[Optimized EG Consumption:] Minimizing excess EG to alleviate large-scale recovery pressures without sacrificing kinetics.
\end{description}

Systematic optimization identified an $\text{EG:PET}$ weight ratio of $5:1$ and a $3\text{-hour}$ reaction time as the critical threshold for complete PET conversion and stable yields ($>85\%$). Furthermore, retaining trace amounts of intermediate oligomers proved beneficial: these solid-phase residues act as natural physical adsorbents during hot-water filtration, effectively trapping residual industrial pigments and reducing the loading of downstream decolorization units.

\subsection{Collaborative Reinforcement (Our New Method)}

\subsection*{Step I: Controlled Chemical Shearing and Degradation in a Twin-Screw Extruder}
To tackle the mass transfer constraints of conventional reactors, a continuous twin-screw extrusion pre-depolymerization stage was introduced to couple thermomechanical forces with chemical agents. As illustrated in intrinsic viscosity ($\text{IV}$) trends (Fig. 1), this phase exhibits distinct kinetic and engineering characteristics:

\vspace{0.3em}
\noindent\textbf{EG-Driven Sharp Degradation and Catalyst-Buffered Control:} 
The initial PET substrate ($\text{IV} \approx 0.687$, average degree of polymerization $\text{DP} \approx 107$) experienced a dramatic $\text{IV}$ drop upon the introduction of trace ethylene glycol (EG). Driven by intense mechanical kneading and electrophilic attack, EG rapidly penetrated and cleaved the polymer main chains. The addition of $1.0\text{ wt}\%$ $\text{Zn(OAc)}_2$ catalyst did not accelerate apparent degradation at very low EG dosages, but instead imparted a smoother, more predictable degradation trajectory, effectively avoiding random thermal fluctuations.

\vspace{0.3em}
\noindent\textbf{Diminishing Marginal Returns and Limitations of Standalone Extrusion:} 
When EG dosage increased to $3.0\text{ wt}\%$, the $\text{IV}$ sharply plummeted to $0.21 \text{--} 0.24$ ($\text{DP} \approx 18$), successfully removing spatial mass transfer barriers. However, further increasing EG to $5.0\text{ wt}\%$ flattened the $\text{IV}$ reduction curve (dropping only to $\approx 0.11$). This demonstrates that within the extremely short residence time of an extruder, relying solely on mechanical chemistry hits a point of diminishing marginal returns, refuting the conventional illusion of achieving complete single-stage deep depolymerization via extrusion alone.

\vspace{0.3em}
\noindent\textbf{Critique of High-Reagent Literature Approaches and Our Mild Breakthrough:} 
While previous studies have extensively explored continuous extrusion depolymerization or high-proportion reagent injection strategies to accelerate PET alcoholysis \cite{Wang2025Controlled, Liu2026Continuous}, closer inspection reveals that such high-reagent modes (often utilizing $1:2$ to $1:4$ EG ratios) suffer from dual traps: (1) Chemical saturation, where excess EG beyond the critical threshold ($3.0\text{ wt}\%$) fails to induce further molecular weight reduction; and (2) Engineering safety hazards, where unreacted free EG (boiling point $196^\circ\text{C}$) undergoes violent flash evaporation under high extrusion temperatures ($\ge 260^\circ\text{C}$), causing pressure spikes, material slippage, and potential explosion risks. Conversely, our $1.0 \text{--} 3.0\text{ wt}\%$ micro-EG strategy anchors the extrusion stage in a "controlled degradation" window. The small-molecule EG is instantly consumed by the PET melt with virtually zero vapor overflow, ensuring safe, stable, and smooth operation.

\vspace{0.3em}
\noindent\textbf{Rheological Advantages for Downstream Integration:} 
At the extrusion endpoint ($\text{IV} \approx 0.2 \text{--} 0.25$, marked by the shaded band in Fig. 1), the material exhibits an optimal waxy/molten state with high fluidity yet sufficient shear resistance to maintain stable die pressure and smooth online pelletizing, providing ideal solid-feedstock precursors for subsequent deep depolymerization.

\vspace{0.5em}
In summary, the core value of extrusion pre-depolymerization lies not in chasing an elusive "one-step completion," but in utilizing precise chemical stoichiometry and thermomechanical coupling to transform macro-macromolecules into highly fluid oligomer slurries. Consequently, the optimized benchmark parameters are established at an $\text{EG:PET}$ weight ratio of $3.0\text{ wt}\%$ and catalyst loading of $1.0\text{ wt}\%$, locking the pre-depolymerized $\text{IV}$ within the golden window of $0.2 \text{--} 0.25$.

\begin{figure}[!t]
    \centering
    \includegraphics[width=\linewidth]{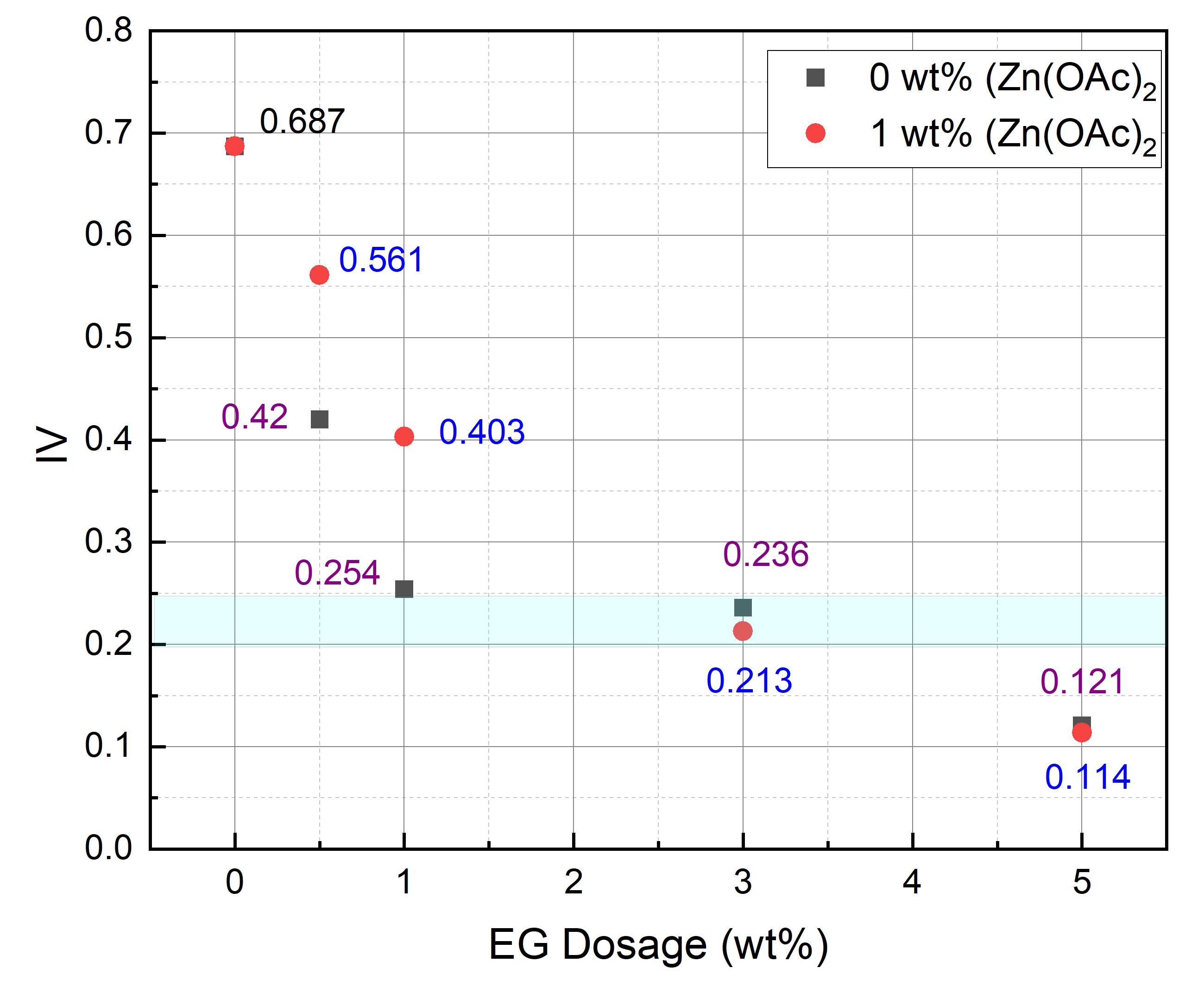}
    \caption{Intrinsic viscosity ($\text{IV}$) of PET as a function of ethylene glycol ($\text{EG}$) dosage under twin-screw extrusion pre-depolymerization with and without $\text{Zn(OAc)}_2$ catalyst (the shaded band highlights the optimized $\text{IV}$ target window of $0.2 \sim 0.25$).}
    \label{fig:iv_trend}
\end{figure}

\subsection*{Step II: Solid-State Comminution and Micro-Homogenization}
Following the twin-screw extrusion stage, which downscales macromolecules into waxy/molten oligomers ($\text{IV} \approx 0.2 \text{--} 0.25$) followed by pelletization, the process seamlessly transitions to the second step: solid-state comminution and micro-homogenization.

Although the pre-depolymerized extrudate exhibits a favorable solid form, short residence times and differential cooling shrinkage can still result in non-uniform macroscopic morphology and micro-segment distribution. To address this, mechanical comminution and homogenization treatments were introduced, delivering several critical engineering functions:

\vspace{0.3em}
\noindent\textbf{Particle Size Refinement and Specific Surface Area Enhancement:} 
Utilizing customized crushing equipment, the pelletized extrudates are further pulverized into $1 \text{--} 2\text{ mm}$ (or finer) particles, dramatically increasing the specific surface area of the solid oligomers.

\vspace{0.3em}
\noindent\textbf{Dual Homogenization of Macro-Morphology and Micro-Composition:} 
While the preceding extrusion step focuses on targeted chemical bond scission (solving the "cleavage" challenge), this step prioritizes absolute physical uniformity (solving the "dispersion and fineness" challenge). Pulverization releases residual internal stresses, ensuring highly consistent molecular weight distribution and physical states across batches.

\vspace{0.3em}
\noindent\textbf{Elimination of Mass Transfer Barriers for Subsequent Deep Depolymerization:} 
These small-sized, highly uniform solid particles enable instantaneous wetting and dead-zone-free solid-liquid contact with catalysts and solvents upon entering the final reactor, completely eradicating mass transfer lags caused by bulky feedstocks and establishing a perfect physical foundation for efficient, complete conversion.

\subsection*{Step III: High-Efficiency Deep Depolymerization and Product Control in the Reactor}
Following the upstream twin-shear pre-depolymerization and solid-state micro-homogenization ($1 \text{--} 2\text{ mm}$ granules), the pre-polymerized oligomers were transferred into a conventional reactor for the main deep alcoholysis stage ($196^\circ\text{C}$, $1.0\text{ wt}\%$ $\text{Zn(OAc)}_2$). Compared to traditional direct one-pot routes, this stage exhibits revolutionary conversion efficiencies and stoichiometric advantages, as corroborated by the BHET yield trends shown in Fig. 2:

\vspace{0.3em}
\noindent\textbf{Overcoming One-Pot Mass Transfer Bottlenecks:} 
In traditional one-pot alcoholysis, large pristine PET particle sizes and high crystallinity impose severe solid-liquid mass transfer barriers, causing BHET yields to plateau at around $85\%$ despite large excess reagent loadings. As illustrated by the asymptotic curve in Fig. 2, increasing the $\text{EG}/\text{Oligomer}$ ratio beyond $2$ yields diminishing returns as the curve flattens.

\vspace{0.3em}
\noindent\textbf{Stoichiometric Efficiency and Yield Enhancement:} 
Empowered by the micro-scale uniform particle size and high specific surface area from preceding steps, the pre-depolymerized oligomers eliminated mass transfer limits. Under a mild $\text{EG}/\text{Oligomer}$ weight ratio of $2$, the BHET yield reliably exceeds $80\%$ ($83\%$), further scaling up to $95\%$ at a ratio of $5$ (Fig. 2).

\vspace{0.3em}
\noindent\textbf{Industrial Balance and "Moderate Retention" Strategy:} 
While an ultra-high yield of $95\%$ proves exceptional chemical efficiency, industrial manufacturing does not strictly require complete monomerization. Moderately retaining a fraction of intermediate oligomers benefits downstream melt decolorization and impurity removal, striking an optimal balance between product purity and processability.

\vspace{0.3em}
\noindent\textbf{Mitigation of Solvent Recovery Burdens:} 
From a macroscopic industrial scale-up perspective, lower EG requirements directly alleviate downstream separation, evaporation, and recovery demands. By eliminating the traditional pathway dependency on massive excess EG, this combined process compresses energy consumption and tail-end recovery costs to a highly economical scale.

\begin{figure}[!t]
    \centering
    \includegraphics[width=\linewidth]{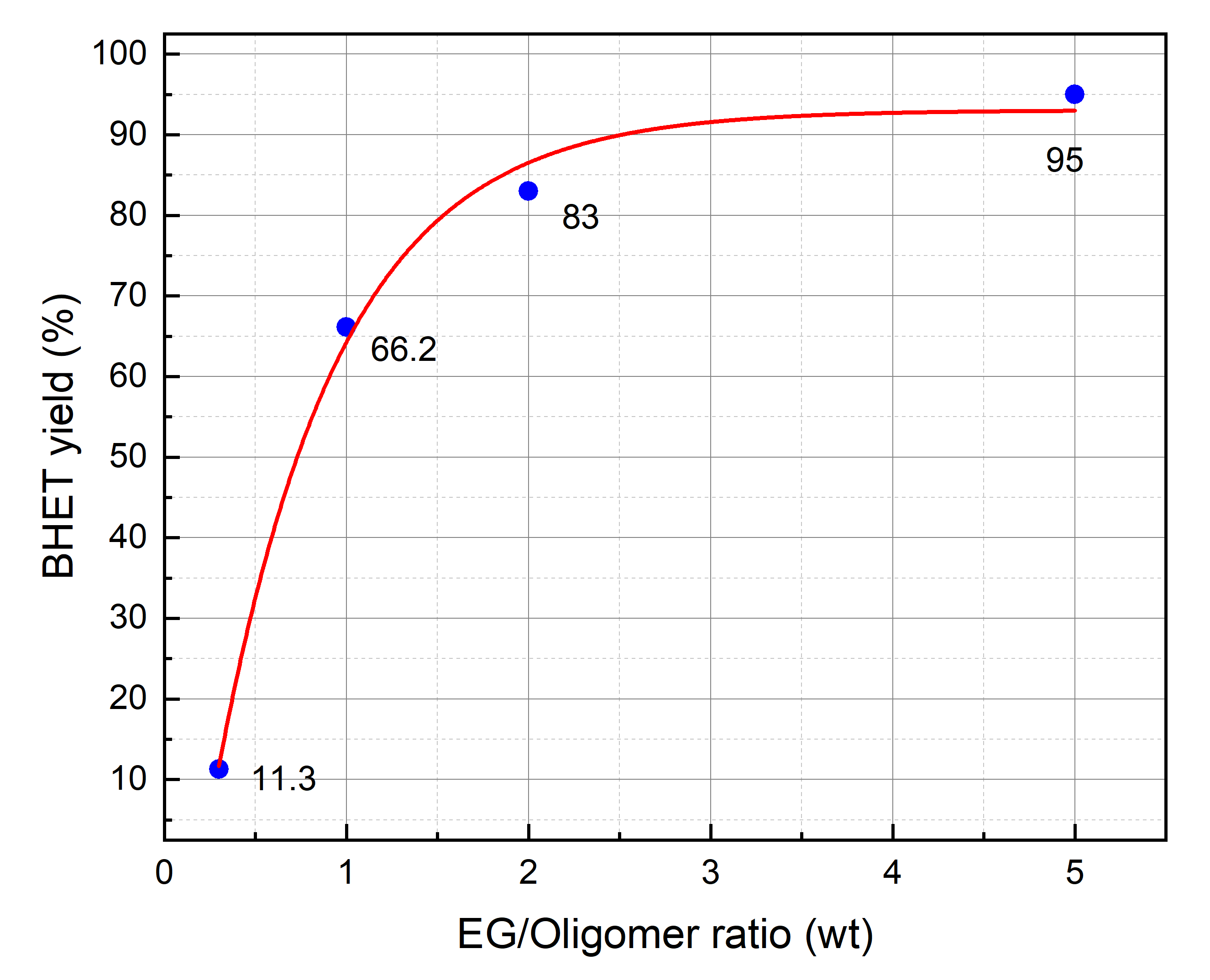}
    \caption{Influence of the $\text{EG/Oligomer}$ weight ratio on the BHET yield during the primary reactor deep depolymerization stage ($196^\circ\text{C}$, $1.0\text{ wt}\%$ $\text{Zn(OAc)}_2$).}
    \label{fig:bhet_yield}
\end{figure}

\vspace{0.3em}
with the $\text{EG/Oligomer}$ weight ratio fixed and optimized at $2$, this study further investigated the kinetic impact of the primary depolymerization reaction time on the BHET yield (Fig. 3). This phase unveils a revolutionary chemical kinetic feature of high industrial value:

\vspace{0.3em}
\noindent\textbf{Ultrafast Kinetics (Achieved Within 10 Minutes):} 
A major drawback of conventional alcoholysis is the requirement for hours of continuous high-temperature reaction. However, within our integrated system featuring upstream extrusion shearing and solid-state micro-homogenization, the primary depolymerization exhibited astonishingly high reaction rates. As shown in Fig. 3, within just $10\text{ minutes}$ of reaction, the BHET yield rapidly reached and stabilized within the high range of $80\% \sim 83\%$ ($80.7\%$).

\vspace{0.3em}
\noindent\textbf{Time-Domain "Immunity" and Eradication of Kinetic Barriers:} 
As the reaction time was further extended from $10\text{ minutes}$ to $30$, $60$, $120$, and $180\text{ minutes}$, the BHET yield curve formed an almost perfect plateau (ranging steadily from $80.1\%$ to $82.8\%$). This robustly proves that by eliminating the mass transfer resistance and crystalline barriers typical of pristine PET, the chemical reaction rate constant of the primary depolymerization was elevated to an exceptional magnitude, achieving thermodynamic equilibrium almost instantaneously.

\vspace{0.3em}
\noindent\textbf{Massive Space-Time Yield (STY) Dividends:} 
This "time-independence" discovery carries milestone industrial significance. It indicates that reaction time is no longer the bottleneck restricting production capacity in continuous or semi-continuous manufacturing. Large reactors traditionally requiring hours can be compressed to residence times of just over ten minutes, triggering an exponential surge in space-time yield (STY) while drastically reducing equipment investment, footprint, and long-cycle high-temperature energy consumption.

\vspace{0.3em}
\noindent\textbf{Balancing Engineering Safety Redundancy with Spatiotemporal Efficiency:} 
Although the main reaction achieves over $80\%$ yield within $10\text{ minutes}$ with negligible yield differences up to $30\text{ minutes}$, considering the robustness required for large-scale industrial continuous operation, the primary reaction time was reliably set at $30\text{ minutes}$. This design leverages ultrafast kinetics while introducing moderate spatiotemporal redundancy to effectively offset potential feed fluctuations, minor thermal gradients, and imperfect mixing during scale-up, ensuring high product uniformity and absolute stability.

\begin{figure}[!t]
    \centering
    \includegraphics[width=\linewidth]{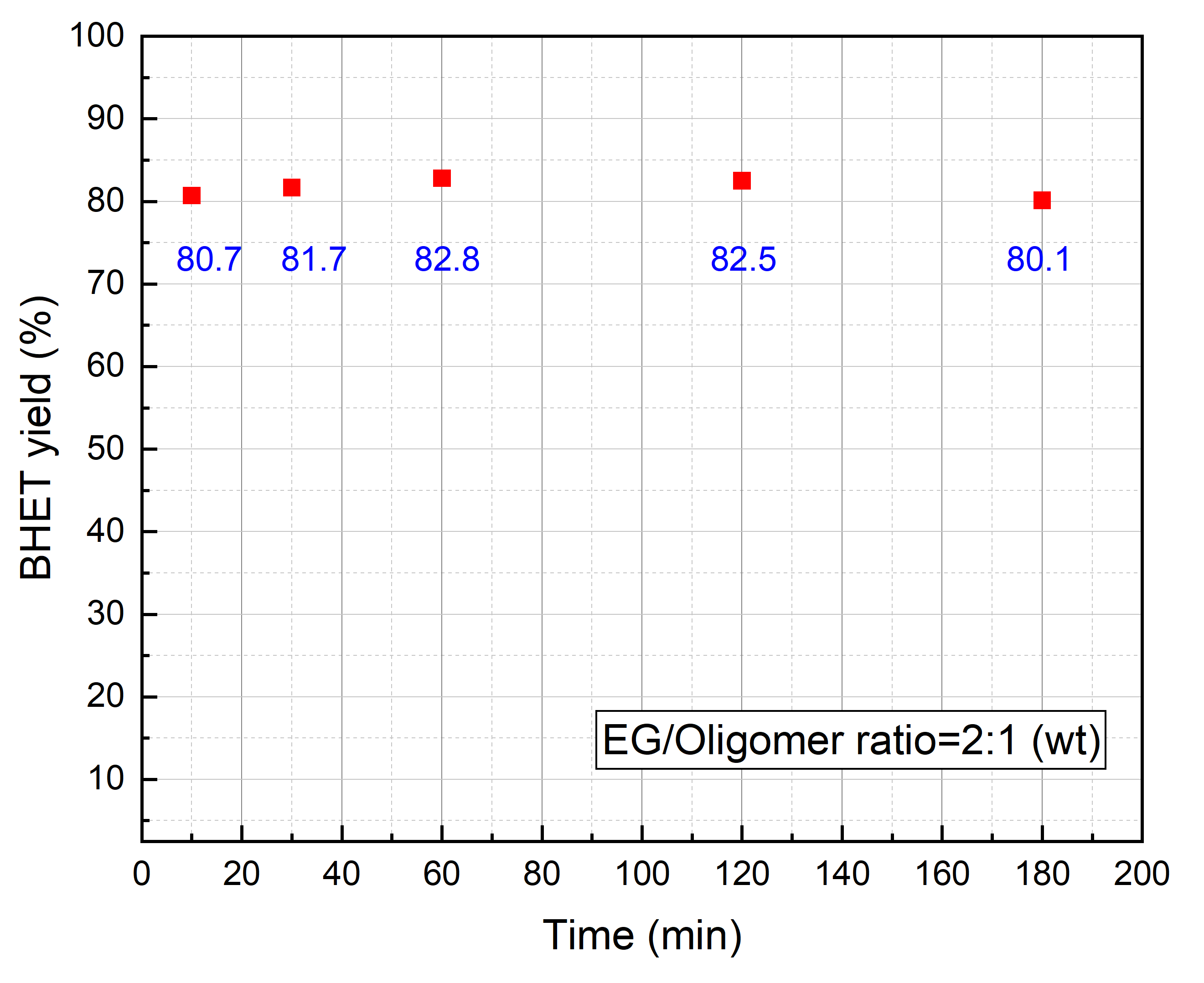}
    \caption{Time-dependent kinetic profile of BHET yield during reactor deep depolymerization at an optimized $\text{EG/Oligomer}$ weight ratio of $2:1$.}
    \label{fig:bhet_kinetic}
\end{figure}

The three-step synergistic reinforcement strategy constructed in this study—Extrusion Controlled Shearing $\rightarrow$ Solid-State Micro-Homogenization $\rightarrow$ Reactor High-Efficiency Deep Depolymerization—thoroughly overcomes the stubborn bottlenecks of poor mass transfer, high reagent consumption, and difficult scale-up inherent in conventional alcoholysis:

\begin{itemize}
    \item \textbf{Step I (Twin-Screw Controlled Shearing):} Introduction of trace $\text{EG}$ ($3.0\text{ wt}\%$) and catalyst rapidly shears macromolecules down to a waxy/molten state ($\text{IV} \approx 0.2 \sim 0.25$), optimizing stoichiometric marginal efficiency while avoiding screw slippage and flash evaporation.
    \item \textbf{Step II (Solid-State Micro-Homogenization):} Fine comminution transforms products into $1 \sim 2\text{ mm}$ granules, exponentially increasing specific surface area and eliminating mass transfer dead zones in downstream reactions.
    \item \textbf{Step III (Reactor High-Efficiency Deep Depolymerization):} Under mild conditions ($\text{EG/Oligomer}$ ratio of $2$, $196^\circ\text{C}$), the reaction exhibits ultrafast kinetics ("achieved within $10$ minutes, safely concluded in $30$ minutes"), stabilizing yields in the $80\% \sim 95\%$ high-efficiency bracket.
\end{itemize}

\noindent\textbf{Conclusion:} Through the coupling of trace reagents and thermomechanical energy, this integrated process achieves efficient, economical, and low-carbon recycling of waste polyester while guaranteeing industrial safety redundancy and significantly alleviating tail-end solvent recovery burdens.

\subsection{Thermodynamic and Kinetic Limits of the Ideal Paradigm}
Before evaluating the practical behavior of waste polyester alcoholysis systems, it is essential to establish a lossless, homogeneous, and mass-transfer-free theoretical limit known as the Zero-Loss Cascade State. Within this theoretical framework, spatial constraints and side reactions are excluded, allowing the collaborative reaction units (SRUs) to form a seamless linear operator product chain.

\vspace{0.5em}
\noindent\textbf{Core Control Equations and Topological Transfer}
The overall monomer production flux of the ideal system, $\Phi_{ideal}(t)$, can be expressed as a continuous operator cascade integral:
\begin{equation}
    \Phi_{ideal}(t) = \left[ \prod_{i \in \{A, B, G, D, E, S\}} \hat{L}_i \right] \cdot \mathcal{S}_0(t)
\end{equation}

\noindent\textbf{Nomenclature \& Operator Definitions:}
\begin{description}
    \item[$\Phi_{ideal}(t)$:] Ideal system monomer production flux ($\text{mol} \cdot \text{s}^{-1}$), representing the instantaneous generation rate of the target monomer over time $t$.
    \item[$\mathcal{S}_0(t)$:] Initial solid-phase potential and geometric distribution term, representing the physical morphology and macroscopic contact substrate of initial PET fragments.
    \item[$\hat{L}_i$:] Transformation operator of the $i$-th synergistic reaction unit ($i \in \{A, B, G, D, E, S\}$), representing the dynamic integral acting on the state vector:
    \begin{itemize}
        \item $\hat{L}_A$ (Penetration Operator): $\int \kappa_A \cdot C_{eff} \, dt$, where $\kappa_A$ is the solid-liquid wetting coefficient and $C_{eff}$ is the effective active ester group density.
        \item $\hat{L}_B$ (Activation Operator): $\exp\left(-\frac{\Delta G^\ddagger_{activation}}{RT}\right) \cdot [Cat]_{clust}$, where $\Delta G^\ddagger_{activation}$ is the coordination activation free energy and $[Cat]_{clust}$ is the effective catalytic site concentration.
        \item $\hat{L}_G$ (Energy Barrier Operator): $k_0 \exp\left(-\frac{E_a}{RT}\right)$, where $k_0$ is the pre-exponential factor, $E_a$ is the apparent activation energy, $R$ is the gas constant, and $T$ is the thermodynamic temperature.
        \item $\hat{L}_D$ (Evolution Operator): $\frac{d}{dt} \text{Dist}(n)$, representing the temporal component spectrum evolution of degree of polymerization $n$.
        \item $\hat{L}_E$ (Competition Operator): $S_1 = \frac{k_1}{k_1+k_2}$, where $S_1$ is the nucleophilic selectivity factor, and $k_1$ and $k_2$ denote the rate constants of the depolymerization pathway and side reactions, respectively.
        \item $\hat{L}_S$ (Calibration Operator): $1 - \delta$, where $\delta$ is the residual defect correction term ($\delta \to 0$ in the ideal state).
    \end{itemize}
\end{description}

\begin{figure}[h]
    \centering
    
    \includegraphics[width=\linewidth]{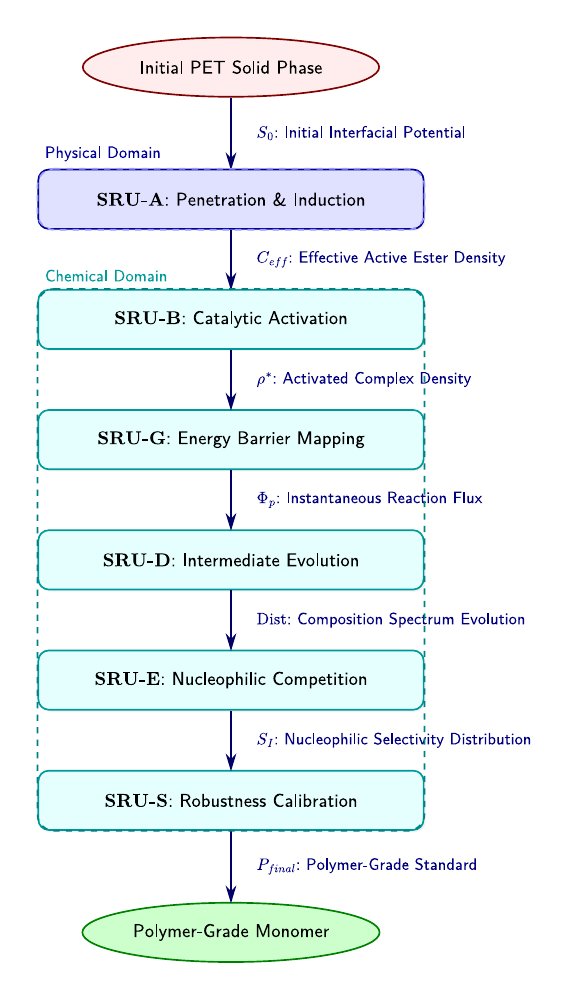}
    \caption{Schematic diagram of the topological operator cascade and synergistic reaction unit (SRU) transfer mechanism in the zero-loss ideal depolymerization state.}
    \label{fig:ideal_cascade}
\end{figure}

In this ideal cascade, every transformation step from the penetration operator ($\hat{L}_A$) to the calibration operator ($\hat{L}_S$) achieves theoretical optimality ($\eta_i = 1$), devoid of energy dissipation or feedback interference ($\Omega_j = 0$).

\vspace{0.5em}
\noindent\textbf{SRU Cascade Mechanism under Ideal States}
Governed by the aforementioned operator chain, the system exhibits a "Monotonic Increasing Principle" ($\frac{\partial \Phi_{ideal}}{\partial \eta_i} > 0$), where each reaction unit transfers mass and energy losslessly at $100\%$ efficiency (as illustrated in Fig. 4):

\begin{description}
    \item[SRU-A ($\to C_{eff}$, Penetration):] Instantaneous solid-liquid interfacial wetting converts macroscopic geometry into maximum-density active contact surfaces.
    \item[SRU-B ($\to \rho^*$, Activation):] Lossless coordination between catalysts and ester groups forms dense activated complexes.
    \item[SRU-G ($\to \Phi_p$, Energy Barrier):] Strict adherence to Arrhenius kinetics enables efficient overcoming of chemical bond energy barriers to output instantaneous flux.
    \item[SRU-D ($\to \text{Dist}$, Evolution):] Polymer chains undergo smooth temporal cleavage, evolving into an ideal oligomer component spectrum.
    \item[SRU-E ($\to S_1$, Competition):] Absolute nucleophilic bias ($S_1 = 1$) is manifested in multi-hydroxyl environments, locking down the deep depolymerization pathway.
    \item[SRU-S ($\to P_{final}$, Calibration):] Minor terminal deviations are instantaneously nullified, directly outputting polymer-grade pure monomers.
\end{description}

\subsection{Physical Reality Deviation and Multi-System Breakdown in Traditional One-Pot Systems}
In realistic one-pot systems, however, this ideal paradigm suffers catastrophic structural breakdown due to two successive layers of physical reality (as illustrated in Fig. 5):

\begin{figure*}[!t]
    \centering
    \includegraphics[width=\textwidth]{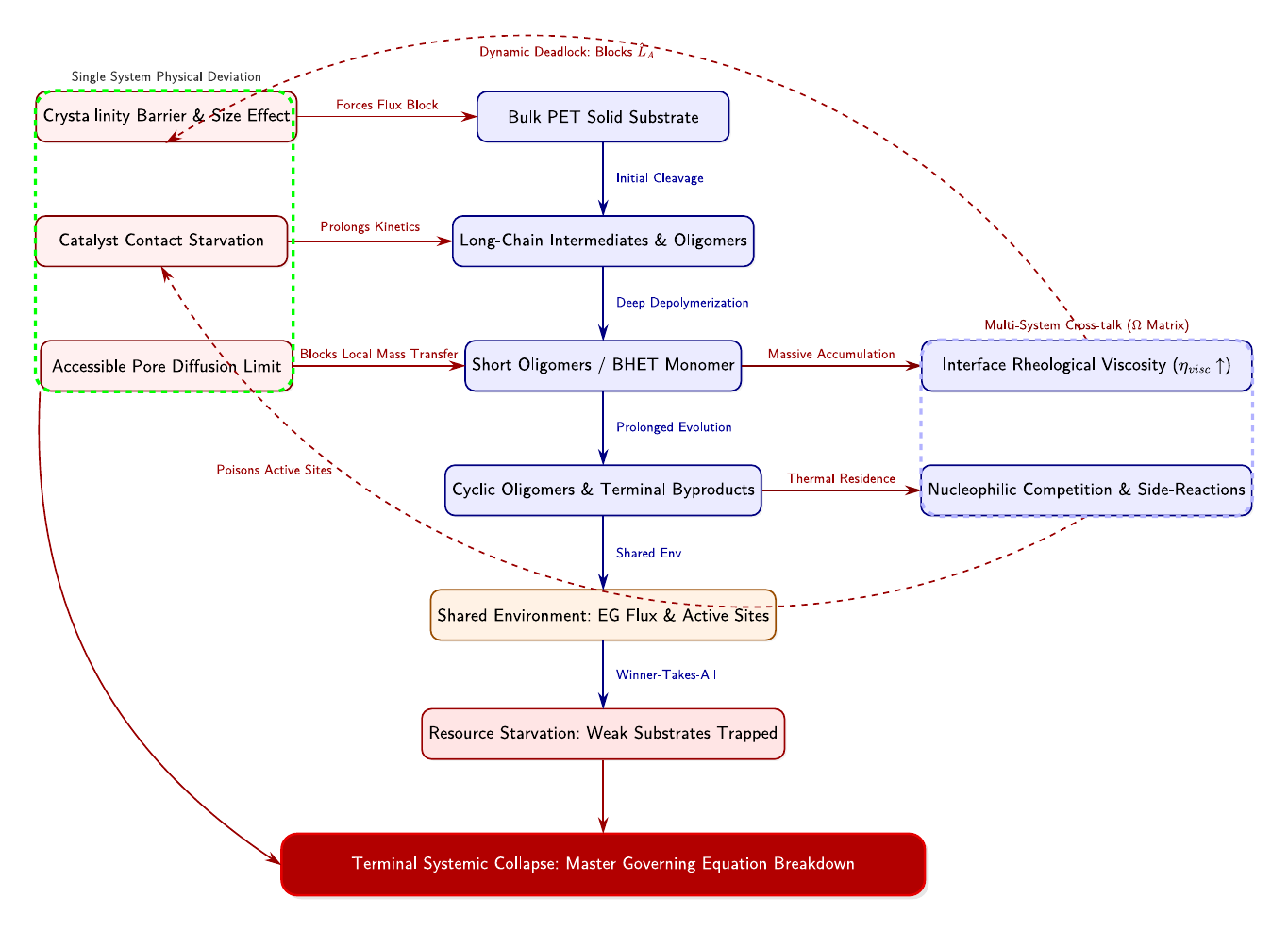}
    \caption{Comprehensive schematic diagram of the master governing equation breakdown and terminal systemic collapse in traditional one-pot depolymerization, illustrating the coupled mechanisms of crystallinity barriers, mass transfer limits, viscosity deadlock, and side-reaction accumulation.}
    \label{fig:master_breakdown}
\end{figure*}

\vspace{0.3em}
\noindent\textbf{Layer I: Single-System Physical Deviations (Size \& Crystallinity Barriers):} 
PET substrates possess macroscopic dimensions and dense crystalline zones. When the $\text{EG/PET}$ ratio drops below critical thresholds, molecular diffusion fails to match surface dissolution rates. This mass transfer resistance ($\hat{L}_A$) forces catalyst contact starvation and stretches chemical kinetics into a protracted multi-hour "waiting effect."

\vspace{0.3em}
\noindent\textbf{Layer II: Multi-System Cross-Talk and Viscosity Deadlock:} 
As depolymerization proceeds, multi-generational intermediates (short-chain oligomers and monomers) trigger severe resource competition. Deep depolymerization products accumulate locally, causing an exponential surge in interfacial rheological viscosity ($\eta_{visc}$). This creates a fatal negative-feedback loop that seals off downstream-to-upstream mass transfer channels while fueling prolonged side-reactions (e.g., DEG formation) that erode yield.

\vspace{0.5em}
Synthesizing these physical deviations and cross-talk feedbacks, the real one-pot system transitions from an ideal linear operator into a complex differential dynamic system governed by a master equation:

\vspace{0.3em}
\noindent\textbf{1. The Master Governing Equation}

\begin{small}
\begin{equation}
\begin{split}
    \Phi_{real}(t) = &\left[ \prod_{i \in \{A \dots S\}} \left( \hat{L}_i^{single} - \sum_{j} \mathcal{H}_{ij}(\Omega) \right) \right] \cdot \mathcal{S}_0(t) \\
    &- \sum_{k} \Gamma_{loss}(t)
\end{split}
\end{equation}
\end{small}

\vspace{0.8em}

Fully expanded in the context of engineering and physical fields, the master equation is expressed with multi-line alignment for clarity:
\begin{align}
    \Phi_{real}(t) &= \left[ \prod \hat{L}_i \right] \cdot \mathcal{S}_0(t) 
    - \underbrace{\vphantom{\int}\kappa_A \cdot \frac{C_{threshold} - C_{eff}(t)}{V_{pore}} \cdot \Theta(t)}_{\text{(1) Mass transfer lag: EG:PET critical collapse}} \nonumber \\
    &\quad - \underbrace{\alpha \cdot \int_{0}^{t} \Phi_D(\tau) \cdot \eta_{visc} \, d\tau}_{\text{(2) Product blocking: Dynamic viscosity deadlock}} \nonumber \\
    &\quad - \underbrace{\beta \cdot \int_{0}^{t_{final}} (1 - S_1) \, dt}_{\text{(3) Kinetic loss: Prolonged side-reaction accumulation}}
\end{align}

\noindent\textbf{Nomenclature \& Physical Meanings:}
\begin{itemize}
    \item \textbf{Macroscopic Evolution Term ($\Phi_{real}(t)$) and Baseline ($\mathcal{S}_0(t)$):}
    \begin{itemize}
        \item $\Phi_{real}(t)$: Actual output or conversion rate matrix of the real one-pot system over time.
        \item $\mathcal{S}_0(t)$: Initial substrate energy and mass baseline under ideal conditions.
    \end{itemize}

    \item \textbf{Layer I: Mass Transfer Lag and Critical Collapse Term ($\Gamma_A$):}
    \begin{itemize}
        \item $\kappa_A$: Effective penetration rate constant at the $\text{SRU}_A$ stage.
        \item $C_{threshold}$ \& $C_{eff}(t)$: Critical solvent concentration threshold required to maintain solid amorphous zone wetting, and the dynamic effective concentration in the reaction system, respectively.
        \item $V_{pore}$: Microscopic pores and accessible volume generated during PET matrix dissolution.
        \item $\Theta(t)$: Heaviside step function. When $C_{eff}(t) < C_{threshold}$ (e.g., when the $\text{EG:PET}$ ratio drops to the $3:1$ critical point), $\Theta(t) \to 1$, mass transfer resistance approaches infinity, and the operator $\hat{L}_A$ completely fails, triggering a cliff-like collapse in conversion.
    \end{itemize}
    \item \textbf{Layer II: Product Blocking and Dynamic Viscosity Deadlock Term ($\Gamma_{D \to A}$):}
    \begin{itemize}
        \item $\alpha$: Negative feedback coupling coefficient of viscosity induced by multi-system product accumulation.
        \item $\Phi_D(\tau)$: Instantaneous concentration of oligomers evolved at the $\text{SRU}_D$ stage within the integration interval $0 \to t$.
        \item $\eta_{visc}$: Interfacial local rheological viscosity. As deep depolymerization products accumulate, $\eta_{visc}$ increases exponentialy, exerting continuous "negative feedback blockage" on frontend penetration via time integration.
    \end{itemize}
    \item \textbf{Layer II: Kinetic Loss and Time Redundancy Term ($\Gamma_E$):}
    \begin{itemize}
        \item $\beta$: Weight damping coefficient of side-reaction pathways.
        \item $t_{final}$: Total redundant reaction time prolonged by the waiting effect in traditional one-pot methods (e.g., $3\text{ hours}$).
        \item $(1 - S_1)$: Deficiency rate of target product selectivity. This term quantitatively evaluates the "$15\%$ yield thermodynamic degradation and byproduct loss caused by waiting" through whole-process time integration.
    \end{itemize}
\end{itemize}

\vspace{0.3em}
\noindent\textbf{2. Ultimate Closed-Loop Explanation of Macroscopic Experimental Phenomena}
When this master governing equation operates at full capacity in reality, it perfectly and irrefutably explains all anomalies observed in experiments:

\noindent\textbf{Why does conversion suffer a cliff-like collapse when $\text{EG:PET} = 3:1$?} \\
Because the step function $\Theta(t)$ in the mass transfer lag term is triggered, rendering the $\hat{L}_A$ operator instantly inactive and resistance infinite, physically "choking off" the system's throat.

\vspace{0.5em}
\noindent\textbf{Why does a simple depolymerization reaction require up to $3\text{ hours}$?} \\
Because multi-system cross-talk and viscosity deadlock ($\eta_{visc}$ integral accumulation) generate massive internal damping, forcing the catalytically active sites into a helpless "waiting effect."

\vspace{0.5em}
\noindent\textbf{Why is there a persistent $15\%$ yield loss?} \\
Because in the absence of spatial or temporal truncation, prolonged thermal residence allows side reactions and competitive pathways ($\Gamma_E$ integral) to wildly devour the target product, ultimately collapsing selectivity.

\subsection{The Integrated Three-Step Reinforcement Approach: Mechanism and Synergy}
As illustrated in Fig. 6, the three-step synergistic reinforcement strategy integrates precise thermomechanical and chemical steps to overcome traditional bottlenecks:

\vspace{0.3em}
\noindent\textbf{Step I: Controlled Chemical Shearing and Degradation in Twin-Screw Extrusion}
As the initiating step of the newly constructed three-step synergistic paradigm, the extrusion stage utilizes a strong coupling of thermomechanical force and trace chemical reagents to achieve "preliminary dimensional reduction" and mass-transfer boundary reconstruction of bulk $\text{PET}$.

\begin{itemize}
    \item \textbf{Physical Reality and Macroscopic Manifestation:} When raw $\text{PET}$ with an initial intrinsic viscosity ($\text{IV}$) of $\approx 0.687$ ($\text{DP} \approx 107$) is fed into the twin-screw extruder, severe degradation occurs within a remarkably short residence time under the synergy of $3.0\text{ wt}\%$ trace $\text{EG}$ and catalyst. The $\text{IV}$ is rapidly suppressed to $0.2 \sim 0.25$ ($\text{DP} \approx 18$), yielding a waxy/molten state possessing both high fluidity and specific shear resistance. This precisely locks onto the marginal efficiency golden inflection point, completely bypassing equipment high-temperature flash evaporation, screw slippage, and shear failure caused by blind high-ratio dosing.
    \item \textbf{Specific Allocation of the Six SRU Operators:} Within this second-level high-temperature/high-shear window, the six operators do not activate indiscriminately; rather, they anchor strictly to frontend physical wall-breaking and initial bond scission:
    \begin{itemize}
        \item $\text{SRU}_A$ (Penetration \& Induction, $\alpha$): $\text{PET}_{(s)} + n\text{Solv} \rightleftharpoons \text{PET}_{(solv-gel)}$. Utilizing intense mechanical kneading and instant dispersion of trace $\text{EG}$, it forcibly breaks the dense crystalline barriers of bulk $\text{PET}$, instantly elevating the effective active ester group density $C_{eff}$ at the solid-liquid interface and eliminating initial mass-transfer resistance.
        \item $\text{SRU}_B$ (Catalytic Activation, $\beta$): $\text{Cat} + [-\text{COO}-] \rightleftharpoons [\text{Cat}\dots\text{O}=\text{C}-\text{O}-]$. Trace catalysts instantaneously construct activated complex density $\rho^*$ in the high-temperature/high-shear microenvironment, imparting controllable reaction buffering capacity.
    \end{itemize}
    \item \textbf{Spatial Isolation:} Deep multi-generational evolution ($\text{SRU}_D$), nucleophilic competition ($\text{SRU}_E$), and robustness calibration ($\text{SRU}_S$) are entirely suspended and unactivated at this stage, cutting off multi-system cross-talk paths at the source.
\end{itemize}

\vspace{0.3em}
\noindent\textbf{Kinetic Control and Governing Equation:}
\begin{equation}
    \frac{d(\text{IV})}{dt} = - k_{extruder} \cdot \sigma_{shear} \cdot C_{EG}^{\alpha} - \Gamma_{SRU-A}
\end{equation}
where $\sigma_{shear}$ represents the high-intensity mechanical shear rate field, and $\Gamma_{SRU-A}$ denotes the mass transfer lag term forced down to near zero by trace reagents and mechanical reinforcement. This equation ensures stable output of quality oligomer granules ($X_{out}$) within a safe zone free of viscosity deadlock.

\vspace{0.5em}
\noindent\textbf{Step II: Solid-State Comminution and Micro-Homogenization}
Following twin-screw extrusion—where macromolecules are downscaled into waxy/molten oligomers ($\text{IV} \approx 0.2 \sim 0.25$) and pelletized—the process naturally transitions to solid-state comminution and micro-homogenization.

\begin{itemize}
    \item \textbf{Physical Reality and Macroscopic Manifestation:} Customized comminution equipment further refines the extrudate into $1 \sim 2\text{ mm}$ (or finer) particles. This precipitous drop in physical dimensions exponentially increases the specific surface area of the oligomer solids, completely releasing internal residual stresses and ensuring high batch-to-batch uniformity in molecular weight distribution and physical state. This eliminates mass transfer lags and reaction non-uniformity caused by bulk feed, establishing a flawless physical foundation for efficient and complete downstream conversion.
    \item \textbf{SRU State Reconstitution (Transition Bridge):} Without direct chemical reaction operator intervention, this stage represents a pure physical morphology reconstruction and operator state transition. Although no chemical bond cleavage occurs, reducing particle size to $1 \sim 2\text{ mm}$ scales up interfacial active ester group density $C_{eff}$, fundamentally clearing mass-transfer dead zones. Other operators ($\text{SRU}_B$, $\text{SRU}_G$, $\text{SRU}_D$, $\text{SRU}_E$, $\text{SRU}_S$) maintain steady-state transition and readiness for the downstream reactor.
    \item \textbf{Kinetic Control:} Pure physical particle-granulation and micro-homogenization control (eliminating the mass transfer resistance term $\Gamma_{SRU-A}$).
\end{itemize}

\vspace{0.5em}
\noindent\textbf{Step III: High-Efficiency Deep Depolymerization and Kinetic Endpoint Advancement}
Following upstream controlled shearing and micro-homogenization, the uniform $1 \sim 2\text{ mm}$ pre-depolymerization particles enter a conventional reactor for deep alcoholysis ($196^\circ\text{C}$, $1\text{ wt}\%$ $\text{Zn(OAc)}_2$ catalyst).

\begin{itemize}
    \item \textbf{Physical Reality and Macroscopic Manifestation:} Leveraging the micrometer/millimeter uniform particle size and ultrahigh specific surface area, the system completely escapes conventional one-pot mass transfer limits. At a moderate $\text{EG/Oligomer}$ weight ratio of $2$, BHET yields rapidly surge and stabilize above $80\%$, scaling up to $95\%$ at a $5:1$ ratio. Furthermore, reaction rate constants are instantaneously boosted, advancing the kinetic endpoint: experiments demonstrate that within just $10\text{ minutes}$, BHET yield reaches and stabilizes in the $80\% \sim 83\%$ high-efficiency bracket. As time extends to $30$, $60$, $120$, and $180\text{ minutes}$, the yield curve forms a perfect horizontal plateau, proving that upstream preprocessing eliminates crystallization barriers and establishes thermodynamic equilibrium instantaneously.
    \item \textbf{SRU Synergy and Endpoint Advancement:} The operator pool transitions from local frontend activation to global efficient synergy. With crystalline barriers and mass-transfer dead zones physically removed, activation energy barriers ($\text{SRU}_G$) and intermediate oligomer evolution ($\text{SRU}_D$) achieve peak product flux $\Phi_p$ within $10\text{ minutes}$. Because upstream reinforcement advances the effective kinetic endpoint to within $10\text{ minutes}$, the subsequent time window exhibits "time-independence." Consequently, the reaction endpoint is scientifically anchored at $30\text{ minutes}$, providing engineering safety redundancy while eliminating prolonged thermal boiling.
\end{itemize}

\vspace{0.3em}
\noindent\textbf{Kinetic Control and Governing Equation:}
\begin{equation}
    \frac{d[\text{BHET}]}{dt} = k_{reaction} \cdot C_{Oligomer} \cdot C_{EG} - \Gamma_{E}
\end{equation}
where elimination of $\Gamma_{SRU-A}$ and suppression of $\Omega$-matrix cross-contamination cause the negative feedback dissipation term $\Gamma_{E}$ to approach zero, driving a magnitude-level surge in reaction rate constants and rapid high-yield breakthroughs.

\begin{figure}[h]
    \centering
    \includegraphics[width=\linewidth]{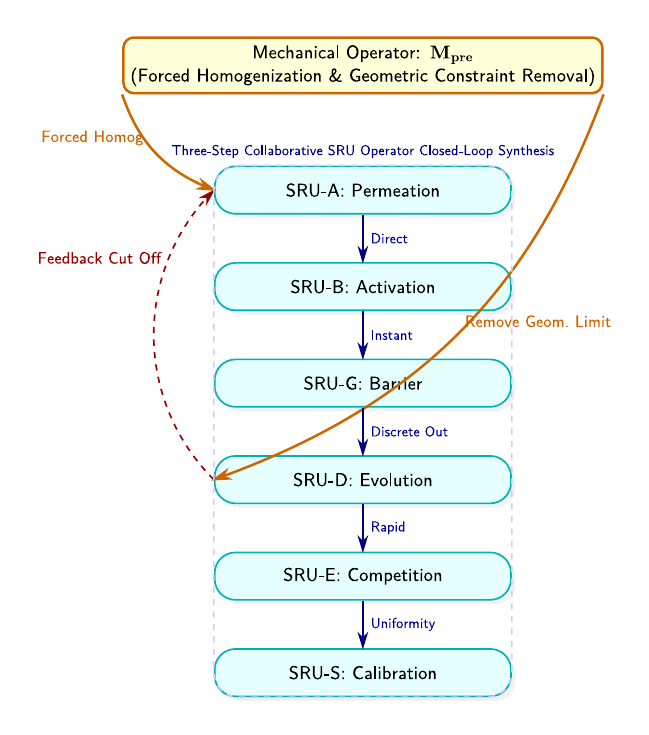}
    \caption{Schematic diagram of the three-step collaborative SRU operator closed-loop synthesis and mechanical intervention mechanism ($\text{M}_{pre}$), illustrating the forced homogenization, geometric constraint removal, and feedback cut-off (\text{FEEDBACK OFF}) that eliminate system collapse.}
    \label{fig:three_step_synergy}
\end{figure}

\section{Conclusions}
In summary, this study resolves a long-standing industrial paradox in $\text{PET}$ chemical recycling: why thousands of high-efficiency laboratory discoveries perpetually fail to bridge the chasm toward industrial scale. By uncovering the core "fault line" of industrial reality—namely, the fatal conflict between uncoupled single-variable optimization and full-process engineering entropy generation—we established a novel human-AI collaborative scientific discovery paradigm grounded in governing reference frames and atomic Scientific Reasoning Units ($\text{SRUs}$).

Through rigorous thermodynamic modeling, topological reduction, and a master governing equation, we successfully quantified the collapse mechanisms of traditional one-pot systems, proving that mass-transfer lag ($\Gamma_A$), dynamic viscosity deadlock ($\eta_{visc}$), and prolonged side-reaction accumulation ($\Gamma_E$) are the true culprits behind high reagent consumption and multi-hour reaction bottlenecks.

To counter these structural breakdowns, we developed an innovative Three-Step Synergistic Reinforcement Paradigm:

\begin{itemize}
    \item \textbf{Twin-Screw Controlled Shearing (Step I):} Utilizing $3.0\text{ wt}\%$ trace $\text{EG}$ and high-shear thermomechanical coupling to precisely anchor the $\text{IV}$ within the golden window of $0.2 \sim 0.25$, successfully bypassing equipment safety hazards and high-temperature flash evaporation.
    \item \textbf{Solid-State Micro-Homogenization (Step II):} Comminuting extrudates into $1 \sim 2\text{ mm}$ uniform granules to exponentially expand specific surface area, eliminating mass-transfer dead zones.
    \item \textbf{High-Efficiency Deep Depolymerization (Step III):} Achieving an ultrafast kinetic breakthrough where $\text{BHET}$ yields surge above $80\%$ within just $10\text{ minutes}$ and safely stabilize at $80\% \sim 95\%$ under a mild $2:1$ $\text{EG}/\text{Oligomer}$ ratio.
\end{itemize}

This integrated architecture transforms $\text{PET}$ glycolysis from a fragile empirical reaction into a self-converging, highly efficient, and low-carbon physical topological system. It eliminates time-domain barriers, yields massive space-time yield ($\text{STY}$) dividends, and drastically reduces tail-end solvent recovery burdens, providing a definitive roadmap for the large-scale industrial closed-loop recycling of waste polyesters.

\bibliographystyle{unsrt} 
\bibliography{references}

\end{document}